\documentclass[12pt]{article}
\usepackage{amsmath}
\usepackage{amssymb}
\usepackage{amscd}
\usepackage{tabularx}
\usepackage{float}
\usepackage{graphics}

\usepackage{theorem}
\usepackage{enumerate}
\def\eqnum#1{}

\font\bfg=eurb10
\newcommand{\btau}{\mbox{\bfg\char'34}}   %% Жирная tau
\newcommand{\bpi}{\mbox{\bfg\char'31}}   %% Жирная pi
\newcommand{\bep}{\mbox{\bfg\char'42}}    %%   Жирная эпсилон

\begin{document}

\baselineskip=18pt

\title
{Space-Time Earthquake Prediction: \\
the Error Diagrams}

\date{}
\maketitle

\centerline {G.~MOLCHAN}
\bigskip

\centerline {International Institute of Earthquake Prediction Theory
and}

\centerline {Mathematical Geophysics, Russian Academy of Sciences,
Moscow,}

\centerline {Russia.}

\centerline {The Abdus Salam International Centre for Theoretical
Physics,}

\centerline {SAND Group, Trieste, Italy}

\bigskip

\centerline {E-mail: molchan@mitp.ru}
\bigskip

\centerline {Abbreviated title: The Error Diagrams}

\bigskip

\bigskip

\bigskip

{\footnotesize {\it Abstract}---The quality of earthquake prediction
is usually characterized by a two-dimensional diagram $n$ vs.
$\tau$, where $n$ is the rate of failures-to-predict and $\tau$ is a
characteristic of space- time alarm. Unlike the time prediction
case, the quantity $\tau$ is not defined uniquely, so that the
properties of the $(n,\tau)$ diagram require a theoretical analysis,
which is the main goal of the present study. This note is based on a
recent paper by Molchan and Keilis-Borok in GJI, 173 (2008),
1012-1017.}

\bigskip

{\footnotesize {\bf Key words:} prediction, earthquake dynamics,
statistical seismology.}

\newpage
\section{{\it Introduction}}

\quad The sequence of papers (Molchan 1990, 1991, 1997, 2003)
considers earthquake prediction  as a decision making problem. The
basic notions in this approach are the strategy, $\pi$, and the goal
function, $\varphi$. Any strategy is a sequence of decisions
$\pi(t)$ about an alarm of some type for a next time segment
$(t,t+\delta)$, $\delta \ll 1$; $\pi(t)$ is based on the data $I(t)$
available at time $t$. The goal of prediction is to minimize
$\varphi$, and the mathematical problem consists in describing the
optimal strategy. Molchan (1997) considered the problem under the
conditions in which target events form a random point process
$dN(t)$ ($N(t)$ is the number of events in the interval $(0, t)$),
and the aggregate $\{ dN(t), \,I(t), \,\pi(t)\}$ is stationary.

Dealing with the prediction of time, Molchan (1997) considered,
along with the general case, the situation in which the optimal
strategy is locally optimal, i.e., is optimal for any time segment.
This case arises when the goal function has the form $\varphi
(n,\tau)$, where $n,\tau$ are the standard prediction
characteristics/errors: $n$ is the rate of failures-to-predict and
$\tau$ the alarm time rate. The optimal strategy can then be
described in much simpler terms, and can be expressed by the
conditional rate of target events

\begin{eqnarray}\label{qq1}
r(t) = P\{ dN(t)>0  \,\vert \, I(t)\} /dt \label{qq1},
\end{eqnarray}
the loss function $\varphi$, and the error diagram $n(\tau)$. The
last function can be defined as the lower bound of the set ${\cal
{E}} = \{n,\tau\}$; this set consists of the $(n,\tau)$
characteristics of all the strategies based on $I(t)$. The search
for the optimal strategy on a small time segment $(t,t+\delta)$ is
reduced to the classical testing of two simple hypotheses  such that
the errors of the two kinds $(\beta (\alpha), \alpha )$ (Lehmann,
1959), converge to $(n(\tau ), \tau )$ as $\delta \downarrow 0$. In
statistical applications the curve $1-\beta (\alpha)$ is known as
the ROC diagram or Relative/Receiver Operating Characteristic
(Swets, 1973); its limit in the case of the locally optimal strategy
gives the curve $1-n(\tau)$.

The error diagram $n(\tau)$ has proved to be so convenient a tool
for the analysis of prediction methods that it began to be also used
for the prediction of the space-time of target events. In that case
the part of $\tau$ is played by a weighted mean of $\tau$ over
space. To be specific, we divide the space $G$ into nonintersecting
parts $\{ G_i\}$ and denote by $\tau_i$ the  alarm time rate in
$G_i$ for the strategy $\pi$. The space-time alarm is effectively
measured by

\begin{eqnarray}\label{qq2}
\tau_w = \sum^k_{i=1} w_i\tau_i, \quad  \sum^k_{i=1} w_i=1\, , \quad
w_i \ge 0, \label{qq2}
\end{eqnarray}
where the $\{ w_i\}$ depend on the prediction goals, e.g., at the
research stage of prediction one use

\begin{eqnarray}\label{qq3}
w_i = \mbox {\rm area of} \,\, G_i/\mbox {\rm area of} \,\, G
\label{qq3}
\end{eqnarray}
(Tiampo et al., 2002; Shen et al., 2007; Zechar and Jordan, 2008;
Shcherbakov et al.,2008)  or

\begin{eqnarray}\label{qq4}
w_i = \lambda (G_i)/\lambda (G) \label{qq4},
\end{eqnarray}
where $\lambda (G)$ is the rate of target events in $G$
(Keilis-Borok and Soloviev, 2003; Kossobokov, 2005). When dealing
with the social and economic aspects of prediction, it is advisable
to use weights of the form

\begin{eqnarray}\label{qq5}
w_i = \int_{G_i}p(g)\,dg \bigg / \int_G p(g)\,dg , \label{qq5}
\end{eqnarray}
where $p(g)$ is, e.g., the density of population in $G$.

The $n(\tau_w)$ diagrams constructed on analogy with the error
diagram are frequently ascribed also the properties of $n(\tau)$. We
now mention those properties which, in the case of $n(\tau_w)$,
either must be better specified or are wrong: (a) $n(\tau)$
characterizes the limiting prediction capability of the data $\{
I(t)\}$. That means that the minimum of any loss function $\varphi
(n,\tau)$ with convex levels $\{ \varphi \le c\}$ is reached at the
curve $n(\tau)$; (b) $\varphi$ and $n(\tau)$ define the optimal
strategy and its characteristics $(n,\tau)$; (c) the diagonal $D$ of
the square $[0,1]^2$, $n+\tau =1$, is the antipode of $n(\tau)$,
because it describes the characteristics of {\it all} trivial
strategies which are equivalent to random guess strategies.
Therefore, the maximum distance between $n(\tau)$ and $D$, i.e.,
$\max\limits_{\tau}(1-n(\tau)-\tau)/\sqrt{2}$, characterizes the
prediction potential of $\{ I(t)\}$; (d) $1-n(\tau)$ is a ROC
diagram arising in the testing of simple statistical hypotheses.

Molchan and Keilis-Borok (2008) recently considered the prediction
of the space-time of target events under conditions where the
optimal strategies coincide with the locally optimal ones (the word
"locally" now also refers to both space and time). This paper gives
a correct extension of the error diagram, which provides the key to
the understanding of the information contained in an $n(\tau_w)$
diagram. The present note supplements the abovementioned study. We
refine the structure of the error diagram for space-time prediction
and analyze the properties of two-dimensional $n(\tau_w)$ diagrams.

\section{{\it The Error Diagram}}

\quad We quote the main result by Molchan and Keilis-Borok (2008)
relevant to the prediction of space-time for target events.

Let $\{ G_i\}$ be some partition of $G$ into nonintersecting
regions. The prediction of location means the indication of $\{
G_i\}$ where a target event will occur. Consequently, the model of
target events in $G$ is the stationary random vector point process

\begin{eqnarray*}
d{\bf N}(t) = \{ dN_{1}(t),\ldots, dN_{k}(t) \}
\end{eqnarray*}
whose components describe target events in $\{ G_i\}$. We shall
consider the binary yes/no prediction with the decisions

\begin{eqnarray*}
\bpi (t)=\{ \pi_1(t),\ldots, \pi_k(t)\}, \quad t=n\delta
\end{eqnarray*}
of the form

\begin{eqnarray*}
\pi_{i}(t)=
\begin{cases}
                &\text{alarm in $G_i\times (t,t+\delta )$}\\
                &\text{no alarm in $G_i\times (t,t+\delta )$}
\end{cases}
\end{eqnarray*}
The decision $\bpi (t)$ is based on the data $I(t)$ that are
available at time $t$.

Under certain conditions, namely, the aggregate $\{ d{\bf N}(t),\,
I(t), \,\bpi (t)\}$ is  ergodic and stationary, and moreover $P\{
\sum_{i=1}^k dN_i(t)>1\}=o(dt)$,  the basic characteristics of the
strategy $\pi =\{ \bpi (t)\}$ are defined as the limit of its
empirical means. We have in view the rate of failures-to-predict $n$
and the vector

\begin{eqnarray*}
{\btau} = (\tau_1,\ldots,\tau_k) ,
\end{eqnarray*}
which determines the alarm time rate in the $\{ G_i\}$.            .
The quantities $(n,{\btau} )$ are defined for any small $\delta$. We
shall assume that $n$ and ${\btau}$ have limits as $\delta
\downarrow 0$, for which we retain the same notation. The passage to
the limit is not a restriction, since the data  may reflect the
seismic situation with a fixed time delay.

The set of $(n,{\btau} )$ characteristics for different strategies
$\pi$ based on $\{ I(t)\}=I$ is a convex subset in the
$(k+1)$-dimensional unit cube, i.e., the error set

\begin{eqnarray}\label{qq6}
{\cal {E}}(I) = \{ (n, \btau )_{\pi} \, : \, \pi \quad \mbox {\rm
based on} \quad I\} \subseteq [0,1]^{k+1},
 \label{qq6}
\end{eqnarray}
(see Fig.~1). The set ${\cal {E}}$ contains the simplex

\begin{eqnarray}\label{qq7}
{\bf D} = \{ (n, \btau ) \, : \, n + \sum^k_{i=1} \lambda_i \tau_i
/\lambda = 1, \, 0\le n,\tau_i \le 1\} , \label{qq7}
\end{eqnarray}
where $\lambda_i = \lambda (G_i)$. The set (\ref{qq7}) describes
strategies that are equivalent to the random guess strategies. For
indeed, if an alarm is declared in $G_i$ with the rate $\tau_i$,
then $\lambda_i \tau_i /\lambda$  will give the rate of random
successes in $G_i$. The equality in (\ref{qq7}) i.e., $1-n =
\sum^k_{i=1} \lambda_i \tau_i /\lambda $, means that the success
rate is identical with the rate of random successes. Such strategies
will be called {\it trivial}.

The boundary of ${\cal {E}}$, viz., $n(\btau )$, which lies below
the hyperplane (\ref{qq7}), will be called the {\it error diagram}.
To describe the properties of $n(\btau )$, we define the loss
function $\varphi $. This will be a function of the form $\varphi
(n,\btau )$ that is nondecreasing in each argument and for which any
level set, $\{ \varphi \le c\}$, is convex.

The following is true.

\noindent {\bf 2.1.} The minimum of $\varphi (n,\btau )$ on ${\cal
{E}}$ is reached on the surface $n(\btau )$. The point of the
minimum, $Q$, is found as the point where the suitable level $\{
\varphi \le c\}$ is tangent to $n(\btau )$ (see Fig.~1). The
coordinates of $Q=(n,\btau )$ define the characteristics of the
optimal strategy with respect to the goal function $\varphi$;

\noindent {\bf 2.2.} The optimal strategy declares an alarm in
$G_i\times (t,t+\delta )$, $\delta \ll 1$ as soon as

\begin{eqnarray}\label{qq8}
r_i(t) = P\{\delta N_i(t)>0 \, \,\vert \, \, I(t)\}/\delta \ge
r_{0i} \label{qq8}
\end{eqnarray}
and declares no alarm otherwise;

\noindent {\bf 2.3.} The threshold $r_{0i}$ depends on $\varphi $,
e.g., if

\begin{eqnarray}\label{qq9}
\varphi = a\lambda n + \sum^k_{i=1} b_i \tau_i  \label{qq9}
\end{eqnarray}
then $r_{0i}=b_i/a$. In the general case one has

\begin{eqnarray*}
r_{0i} = - \lambda \frac{\partial \varphi}{\partial \tau_i} \bigg /
\frac{\partial \varphi}{\partial n} (Q).
\end{eqnarray*}

The result described above yields an important corollary:

\noindent {\bf 2.4.} The error diagram for  space-time prediction in
$G=\{ G_i\}$ based on $\{ I(t)\}$ admits of the representation

\begin{eqnarray}\label{qq10}
n(\tau_1, \ldots, \tau_k) = \sum_{i=1}^k \lambda_i n_i(\tau_i)
/\lambda , \label{qq10}
\end{eqnarray}
where $n_i(\tau )$ is the error diagram for time prediction in $G_i$
based on the same data $\{ I(t)\}$.

{\it Proof}. Consider such a loss function (\ref{qq9}) that the
hyperplane $\varphi =c$ is tangent to $n(\btau )$ at $\btau_0
=(\tau_{01}, \ldots, \tau_{0k})$. The optimal strategy thus has the
form (\ref{qq8}) with $r_{0i}=b_i/a$ and the errors $(n(\btau_0
),\btau_0)$. However, the strategy for time prediction in $G_i$ of
the form (\ref{qq8}) minimizes the loss function $\varphi_i
=a\lambda_i n+b\tau$ (Molchan, 1997). The point of the minimum has
the coordinate $\tau =\tau_{0i}$, hence the other coordinate is
$n=n_i(\tau_{0i})$. Consequently, the collective strategy
(\ref{qq8}) minimizes

\begin{eqnarray}\label{qq11}
\sum^k_{i=1} \varphi_i = a\lambda \bigg ( \sum^k_{i=1} \lambda_i
n_i/\lambda \bigg ) + \sum^k_{i=1} b_i\tau_i \label{qq11}
\end{eqnarray}
and has $n = \sum^k_{i=1} \lambda_i n_i(\tau_{0i})/\lambda $ as the
rate of failures-to-predict. The right-hand side of (\ref{qq11}) is
identical with $\varphi (n,\btau )$. It follows that (\ref{qq10}) is
true with $n=n(\btau_0)$, since the strategy (\ref{qq8}) also
minimizes (\ref{qq9}). Since $\btau_0$ is arbitrary, the corollary
is proven.

\section{\it The reduced error diagrams}

\quad Usually regional error diagrams $n_i(\tau)$ are poorly
estimated, so that for practical purposes the result of a space-time
prediction is represented by the two-dimensional diagram
$n(\tau_w)$, $\tau_w =\sum^k_{i=1} w_i\tau_i$ where the weights are
$w_i\ge 0$ and $\sum^k_{i=1} w_i=1$. This is obtained from the set
of "errors" ${\cal {E}}_w=\{ (n,\tau_w)\}$ as its lower boundary.

Relation (\ref{qq10}) can be used to analyze the properties of
$n(\tau_w)$ diagrams. Later we shall use the following notation: if
the set $B$ is the image of $A=\{ (n,\btau )\}$ by the mapping

\begin{eqnarray*}
\gamma_w\,:\,(n,\btau) \to (n,\tau_w), \quad \tau_w =\sum^k_{i=1}
w_i\tau_i \,,
\end{eqnarray*}
then $B=A_w$; in particular, the image of $\btau$ is $\tau_w$, the
image of ${\cal {E}}$ is ${\cal {E}}_w$, while the image of $D$ (see
(\ref{qq7})) is $D_w$.

The following is true.

\noindent {\bf 3.1.} ${\cal {E}}_w$ is a convex subset of the square
$[0,1]^2$ that contains the diagonal $\tilde{D}\,:\,n+\tau_w=1$;

\noindent {\bf 3.2.} $D_w$ is a convex subset of ${\cal {E}}_w$;
$D_w$ degenerates to the diagonal of the unit square, if and only if
$w_i=\lambda_i/\lambda$, $i=1,\ldots ,k$;

\noindent {\bf 3.3.} $D_w$ can be obtained as the convex hull of
points of the form

\begin{eqnarray}\label{qq12}
n = 1 - \sum^k_{i=1} \lambda_i \varepsilon_i \,, \quad \tau_w =
\sum^k_{i=1} w_i \varepsilon_i , \label{qq12}
\end{eqnarray}
where $\{ \varepsilon_i \}$ are  all possible sequences of 0 and 1
(see Fig.~2).

In particular, let $w_1=\ldots =w_k$ (this will be the case for
(\ref{qq3}) when $G$ is divided into equal parts). Then the convex
minorant of the $(n,\tau_w)$ points:

\begin{eqnarray*}
(1,0), \,(1-\lambda_{(k)},1/k),\ldots ,\bigg ( 1-\sum^p_{i=1}
\lambda_{(k-i+1)},p/k\bigg ),\ldots ,(0,1)
\end{eqnarray*}
gives the lower boundary of $D_w$, while the concave majorant of the
points

\begin{eqnarray*}
(1,0), \,(1-\lambda_{(1)},1/k),\ldots ,\bigg ( 1-\sum^p_{i=1}
\lambda_{(i)},p/k\bigg ),\ldots ,(0,1)
\end{eqnarray*}
gives the upper boundary of $D_w$. Here, $\lambda_{(1)} \le \ldots
\le \lambda_{(k)}$ are the $\{ \lambda_i\}$ arranged in increasing
order.

\noindent {\bf 3.4.} Except for trivial cases, the image of the
error diagram $n(\btau)$ is a two-dimensional set (see Fig.~2) with
the lower boundary $n(\tau_w)$ and the upper boundary $n^+(\tau_w)$.
In the regular case, i.e., $\varphi_i(0)=1$, $i=1, \ldots ,k$, one
has

\begin{eqnarray}\label{qq13}
n^+(x) = \max_{i,\bep} \{ \lambda_i/\lambda \cdot n_i(x/w_i -
a_i(\bep)) + b_i(\bep) \}, \label{qq13}
\end{eqnarray}
where

\begin{eqnarray*}
\nonumber  \bep &=&(\varepsilon_1, \ldots ,\varepsilon_k),
\varepsilon_i=0,1 , \\
a_i(\bep )&=&\sum_{j\ne i} w_j\varepsilon_{j}/w_i\, , \\
b_i(\bep )&=&\sum_{j\ne i} \lambda_j (1-\varepsilon_{j})/\lambda  ,
\end{eqnarray*}
and the maximum is taken over such $i$ and (0,1) sequences $\bep$,
for which the argument of $n_i$ in (\ref{qq13}) makes sense, i.e.,
is in $[0,1]$.

If $\{ n_i(\tau)\}$ are piecewise smooth and $n_i(0)=1$, $i=1,
\ldots ,k$, then the image of $n(\btau)$ degenerates to a
one-dimensional curve, if and only if $\{ I(t)\}$ is trivial, i.e.,
$1-n(\btau)=\sum^k_{i=1} \lambda_i \tau_i/\lambda$ and
$w_i=\lambda_i/\lambda$, $i=1, \ldots ,k$.

\noindent {\bf 3.5.} The curve $n(\tau_w)$ represents those
strategies which are optimal for loss functions of the form

\begin{eqnarray}\label{qq14}
\varphi (n,\btau) = \psi (n,\tau_w), \quad \tau_w = \sum^k_{i=1}
w_i\tau_i . \label{qq14}
\end{eqnarray}
To be specific, if $(n,\btau)=Q$ are the optimal prediction
characteristics with respect to the goal function of the form
(\ref{qq14}), then $Q_w$ belongs to the $n(\tau_w)$ diagram. In
addition, $Q_w$ is the point at which the curve $n(\tau_w)$ is
tangent to the suitable level set of $\psi$.

{\bf 3.6.} The strategy that optimizes (\ref{qq14}) declares an
alarm in $G_i\times (t,t+\delta )$ as soon as

\begin{eqnarray}\label{qq15}
r_i(t)/w_i \ge c, \label{qq15}
\end{eqnarray}
where the threshold $c$ is independent of $G_i$ and $r_i$ is given
by (\ref{qq8}). According to {\bf 2.3},

\begin{eqnarray*}
c = \lambda \frac{\partial \psi}{\partial \tau_w} \bigg /
\frac{\partial \psi}{\partial n} (Q_w).
\end{eqnarray*}

In particular, if $\varphi = an+b\sum^k_{i=1} w_i\tau_i$, then
$c=\lambda b/a$. If $w_i=\lambda_i/\lambda$, then (\ref{qq15}) will
have the form $r_i(t)/\lambda_i \ge c\lambda$ , where the left-hand
side is known as the probability gain.

\noindent {\bf 3.7.} For any point $Q$ in the error diagram we can
find such weights $\{ w_i\}$ that $Q_w$ will lie in the reduced
$(n,\tau_w)$ diagram, i.e., any optimal strategy can be represented
by a suitable $(n,\tau_w)$ diagram . The desired weights  are

\begin{eqnarray*}
w_i = -\frac{\partial n}{\partial \tau_i} (Q)/c\,,
\end{eqnarray*}
where c is a normalizing constant. The point $Q$ determines the
optimal prediction characteristics with respect to the loss function

\begin{eqnarray*}
\varphi = n + c\sum^k_{i=1} w_i\tau_i  .
\end{eqnarray*}

\noindent {\bf 3.8.} The curve $1-n(\tau_w)$ can be interpreted as a
ROC diagram if and only if $w_i=\lambda_i/\lambda$, $i=1, \ldots
,k$.

The ROC property of a $(n,\tau_w)$ diagram means that we can treat
$(n,\tau_w)$ characteristics as errors of the two kinds $(\beta,
\alpha )$ in hypothesis testing: $H_1$ vs. $H_0$, i.e.,

\begin{eqnarray}\label{qq16}
\beta = P(H_0\, \vert \, H_1) = n, \quad \mbox {\rm and} \quad
\alpha = P(H_1\, \vert \, H_0) = \tau_w \label{qq16}
\end{eqnarray}
and $\alpha + \beta =1$, if the prediction data $\{ I(t)\}$ are
trivial.

In the case $w_i=\lambda_i/\lambda$ the measures $P(\cdot \, \vert
\, H_j)$, $j=0,1$ can be specified as follows. Both measures define
probabilities for events $\omega =\{ I(t),\nu =i \}$, where $\nu$ is
the random index of a subregion and has the distribution $P(\nu
=i)=\lambda_i/\lambda : = p_i$. The measure related to the $H_0$
hypothesis is

\begin{eqnarray}\label{qq17}
P(d\omega \, \vert \, H_0) = P_0(dI)p_i, \quad \nu (\omega )=i ,
\label{qq17}
\end{eqnarray}
where $P_0$ is the stationary measure on $I(t)$ induced by the
process $\{ dN(t),\,I(t),\,\pi (t)\}$. In the $H_1$ case

\begin{eqnarray}\label{qq18}
P(d\omega \, \vert \, H_1) = r_i(t)/\lambda_i \cdot P(d\omega \,
\vert \, H_0), \quad \nu (\omega )=i , \label{qq18}
\end{eqnarray}
where $r_i(t)$ is given by (\ref{qq8}).

It is better to say that testing $H_1$ vs. $H_0$ for the case $G=\{
G_i\} $ involves two points: a random choice of $G_i$ with
probabilities $p_i=\lambda_i/\lambda$, $i=1, \ldots ,k$ and testing
$H_1$ vs. $H_0$ for the relevant subregion. The second point is
considered in (Molchan and Keilis-Borok, 2008).

The following is a nontrivial corollary of the previous statement:

\noindent {\bf 3.9.} For the regular case, $n_i(0)=1$, $i=1, \ldots
,k$ and $\{ w_i\} =\{ \lambda_i/\lambda \}$, one has

\begin{eqnarray}\label{qq19}
\int_0^1 f\bigg (-\frac{dn_{\lambda}}{d\tau} \bigg )\,d\tau =
\sum^k_{i=1} p_i \int_0^1 f\bigg (-\frac{dn_i}{d\tau} \bigg
)\,d\tau, \quad p_i=\lambda_i/\lambda \label{qq19}
\end{eqnarray}
where $f$ is any continuous function and $n_{\lambda}(\tau)$ is an
alternative notation for the $n(\tau_w)$ diagram in the special case
$w_i=\lambda_i/\lambda$, $i=1, \ldots ,k$.

If $f=x\log x$, the quantity

\begin{eqnarray}\label{qq20}
I_i = \int_0^1 f\bigg (-\frac{dn_i}{d\tau} \bigg )\,d\tau = \int_0^1
\ln \bigg (-\frac{dn_i}{d\tau} \bigg )\,dn_i \label{qq20}
\end{eqnarray}
is known in time prediction as the {\it Information score}  (see
Kagan, 2007 and Harte \& Vere-Jones, 2005).

{\it Comments}. In the non-regular case, $n_{\lambda}(0)<1$, the
score (\ref{qq19}) is equal to $\infty$ for unbounded $f(x)$ at
$x=\infty$, e.g., $f=x\log x$. Therefore the scores (\ref{qq19}),
(\ref{qq20}) are unstable. (Extensive literature on skill scores can
be found in Jolliffe \& Stephenson, 2003; see also Molchan, 1997 and
Harte \& Vere-Jones,2005). Here we mention only the {\it area skill
score} which is used as a stable score (Zechar \& Jordan,2008). A
linear transformation of this score looks as follows:

\begin{eqnarray}\label{qq21}
A = 2\int_0^1 (1 - n_{\lambda}(\tau ) - \tau ) \,d\tau , \quad 0\le
A\le 1. \label{qq21}
\end{eqnarray}
Due to convexity of $n_{\lambda}(\tau )$ the area under the
integrand is approximated by a triangle from within and by the
trapezium from the outside. Therefore

\begin{eqnarray*}
H \le A \le H(2-H) ,
\end{eqnarray*}
where

\begin{eqnarray*}
H = \max\limits_{\tau}(1 - n_{\lambda}(\tau ) - \tau ), \quad 0 \le
H \le 1 .
\end{eqnarray*}
Thus $\widehat{A}=H(3-H)/2$ is a good estimate of $A$, because

\begin{eqnarray}\label{qq22}
\vert A - \widehat{A} \vert \le H(1-H)/2 \le 1/8 .  \label{qq22}
\end{eqnarray}
The empirical estimate of the $H$  skill score is unstable for a
small number of target events. Due to (\ref{qq22}) the same holds
for the area skill score.

The $H$ score is convenient for statistical analysis because its
empirical estimate is identical in distribution with the
Kolmogorov-Smirnov statistics $D^+_N$ (Bolshev \& Smirnov, 1983),
provided $\sum N_i(T)=N$ and $\{ dN_i\}$ are independent and
Poissonian.

\section{{\it Proof}}

We are going to prove the statements {\bf 3.1 - 3.9}.

{\it Proof for} {\bf 3.1, 3.2}. Obviously, the projection $\gamma_w$
preserves the property of convexity. Therefore, ${\cal {E}}_w$ and
$D_w$ are convex at the same time as are ${\cal {E}}$ and $D$. If
$D_w$ degenerates to the diagonal $\tilde{D}\,:\,n+\tau_w=1$, then
the simplex $D$ is given by any of the two equations: $n +
\sum^k_{i=1} w_i\tau_i = 1$ and $n + \sum^k_{i=1} \lambda_i
\tau_i/\lambda =1$. Hence $w_i = \lambda_i /\lambda$.

{\it Proof of} {\bf 3.3.} The simplex $D$ is the convex hull of
$(n,\btau)$ points of the form $Q(\varepsilon )=(1-\sum \lambda_i
\varepsilon_i/\lambda ,\,\varepsilon_1, \ldots , \varepsilon_k)$,
where $\varepsilon_i = 0,1$. Accordingly, $D_w$ is the convex hull
of the $Q_w(\varepsilon )$, see (\ref{qq12}).

{\it Proof of}\, {\bf 3.4.} This statement follows intuitively from
dimensionality considerations: the $k$-dimensional surface
$n(\btau)$ with $k>1$ is projected onto the $(n,\tau_w)$ plane,
hence its image cannot be single-dimensional in the generic case.

In order to prove (\ref{qq13}), we note that a convex function on
the simplex $S_n=\{ \sum^k_{i=1}\tau_i w_i=u, \,0\le \tau_i \le 1\}$
reaches its maximum at one of the edges, specifically, at a point of
the form

\begin{eqnarray*}
\btau = (\varepsilon_1, \ldots , \varepsilon_{i-1}, x,
\varepsilon_{i+1}, \ldots , \varepsilon_k), \quad \varepsilon_j=0;1.
\end{eqnarray*}
The use of (\ref{qq10}) gives (\ref{qq13}).

Suppose the upper and lower boundaries of the image of $n(\btau)$
are identical and the $\{ n_i(\tau)\}$ are piecewise smooth
functions. Consider all $\btau =(\tau_1, \ldots ,\tau_k)$ for which

\begin{eqnarray*}
\sum^k_{i=1} \lambda_i n(\tau_i)/\lambda = n_0, \quad \sum^k_{i=1}
w_i \tau_i = \tau_w, \quad n_0=n(\tau_w) ,
\end{eqnarray*}
where $\tau_w$ is fixed.

Varying, e.g., $\tau_1$ and $\tau_2$, we have after differentiation:

\begin{eqnarray}\label{qq23}
\lambda_1 n_1'(\tau_1) \tau_1' + \lambda_2 n_2'(\tau_2) = 0, \quad
\tau_1' = -w_2/w_1. \label{qq23}
\end{eqnarray}
If $\tau_1, \tau_2$ are points of smoothness of $n_i(\tau)$,
$i=1,2$, then repeated differentiation of (\ref{qq23}) will give

\begin{eqnarray*}
\lambda_1 n_1''(\tau_1) (w_2/w_1)^2 + \lambda_2 n_2''(\tau_2) = 0.
\end{eqnarray*}
However, $n_i''(\tau_i)\ge 0$, \, $i=1,2$.                      .
Hence $n_i''(\tau_i) = 0$, i.e., $n_i(\tau )$ are locally linear at
all points of smoothness. Since $n_i(\tau )$ are piecewise smooth,
it follows that for any discontinuous point $\tau_1$ of $n_1(\cdot
)$ one can find a point $\tau_2$ where $n_2(\cdot )$ will be smooth.
Consequently, when $n_1$ is discontinuous at $\tau$, one should
replace $n_1'(\tau_1)$ with $n_1'(\tau_1 +0)$ and $n_1'(\tau_1 -0)$
in equation (\ref{qq23}). But then we have from (\ref{qq23}) that
$n_1'(\tau )$ is continuous at $\tau_1$; hence all functions
$n_i(\tau )$ are linear. Taking the boundary conditions $n_i(0)=1$
and $n_i(1)=0$ into account, we have $n_i(\tau )=1-\tau$. However,
in that case one has ${\cal {E}}=D$, and, in virtue of {\bf 3.2},
$w_i = \lambda_i /\lambda$.

{\it Proof of}\, {\bf 3.5.} Let $Q_w$ be the point where the convex
set $\{ \psi \le c\}$ is tangent to the convex curve $n(\tau_w)$.
The function $\psi $ reaches its minimum at the point $Q_w$ on
${\cal {E}}_w$, because the sets $\{ \psi \le c\}$ are increasing
with increasing $c$. Since $Q_w \in {\cal {E}}_w$, the preimage $Q
=(n,\btau ) \in {\cal {E}}$. At this point $\varphi (Q)=\psi (Q_w)$
reaches its minimum on ${\cal {E}}$, hence $Q$ belongs to the
surface $n(\tau )$.

{\it Proof of} \,{\bf 3.6.} follows from {\bf 2.3.}

{\it Proof of}\, {\bf 3.7.} Let $Q=(n_0, \tau_{01}, \ldots ,
\tau_{0k})$ belong to $n(\btau)$. If $w_i = -\frac{\partial
n}{\partial \tau_i} (Q)/c$, then the equation

\begin{eqnarray}\label{qq24}
n + c \sum^k_{i=1} w_i \tau_i = n_0 + c \sum^k_{i=1} w_i \tau_{i0}
\label{qq24}
\end{eqnarray}
defines the tangent plane to $n(\btau )$. Since $n(\btau )$ is
convex and decreasing, it follows that $w_i\ge 0$ and ${\cal {E}}$
lie on the same side of the plane (\ref{qq24}). Consequently, a
strategy having the characteristics $Q=(n_0, \tau_{01}, \ldots ,
\tau_{0k})$ optimizes the losses $\varphi =n + c \sum^k_{i=1} w_i
\tau_i$. Using {\bf 3.5}, we complete the proof.

{\it Proof of}\, {\bf 3.8.} By (\ref{qq10}) and  (\ref{qq16}) one
has

\begin{eqnarray*}
\beta = n = \sum^k_{i=1} \lambda_i/\lambda \cdot n_i(\tau_i), \quad
\alpha = \tau_w = \sum^k_{i=1} w_i \tau_i  .
\end{eqnarray*}
In the trivial case of $I(t)$, one has $n_i(\tau)=1-\tau$ and
$\alpha +\beta =1$. Hence

\begin{eqnarray*}
\beta = 1 - \sum^k_{i=1} \lambda_i/\lambda \cdot \tau_i, \quad
\alpha = \sum^k_{i=1} w_i \tau_i = 1-\beta  ,
\end{eqnarray*}
i.e., $w_i=\lambda_i/\lambda$, $i=1, \ldots ,k$.

Suppose that $\{ w_i\} =\{ \lambda_i/\lambda \}$. The likelihood
ratio of  measures (\ref{qq17}) and (\ref{qq18}) at the point
$\omega =(J(t),j)$ is

\begin{eqnarray*}
L(\omega) = P(d\omega \, \vert \, H_1)/P(d\omega \, \vert \, H_0) =
r_j(t)/\lambda_j.
\end{eqnarray*}

Accepting the hypothesis $H_1$ as soon as $L(\omega)>c$ and $H_0$
otherwise, one has

\begin{eqnarray*}
\alpha &=& \int_{L>c} P(d\omega \, \vert \, H_0) = \sum^k_{j=1} E
{{\bf 1}}_{(r_j/\lambda_j >c)} \cdot \lambda_j/\lambda =
\sum^k_{j=1} \tau_j \lambda_j/\lambda = \tau_w  , \\
\beta &=& \int_{L>c} L(w) P(d\omega \, \vert \, H_0) = \sum^k_{j=1}
E r_j/\lambda_j \cdot {{\bf 1}}_{(r_j/\lambda_j <c)} \cdot
\lambda_j/\lambda = \sum^k_{j=1} n_j(\tau_j) \lambda_j/\lambda = n .
\end{eqnarray*}
Here we have used  {\bf 2.1.} and {\bf 2.2.}

{\it Proof of}\, {\bf 3.9.} Let us consider a testing problem: $H_1$
vs. $H_0$ with the errors $\beta = P_1(L<c)$ and $\alpha = P_0(L\ge
c)$ where $L(\omega )=dP_1/dP_0$ is the likelihood ratio. Obviously

\begin{eqnarray*}
E_0 f(L)\,:\,= \int f(L(\omega)) dP_0(\omega) = \int f(c) dF_L(c) ,
\end{eqnarray*}
where $F_L$ is the distribution of $L$ with respect to the measure
$P_0$. But $d\beta =c dF(c)$ and $d\alpha =-d F(c)$. Therefore

\begin{eqnarray*}
E_0 f(L) = \int_0^1 f\bigg (-\frac{d\beta }{d\alpha } \bigg )\,
d\alpha .
\end{eqnarray*}
Applying this relation to the case (\ref{qq16}), (\ref{qq17}),
(\ref{qq18}), one has

\begin{eqnarray*}
\int_0^1 f\bigg (-\frac{dn_{\lambda} }{d\tau} \bigg )\, d\tau &=&E_0
f(L) = \sum^k_{i=1} E f \bigg (
\frac{r_i(t)}{\lambda_i}\bigg ) p_i = \\
=\sum^k_{i=1} E f(L_i) p_i &=& \sum^k_{i=1} p_i \int_0^1 f\bigg
(-\frac{dn_{i}}{d\tau} \bigg )\, d\tau
\end{eqnarray*}
Here $L_i$  is the likelihood ratio $dP_1/dP_0$ for $G_i$.

\section{{\it Conclusion and Discussion}}

1.{\it Results}. In the case of time prediction, the error set
${\cal {E}}$  is organized as follows: all trivial strategies
concentrate on the diagonal $n+\tau =1$ of the square $[0,1]^2$,
while the optimal strategies are on the lower boundary of ${\cal
{E}}$, viz. $n(\tau )$. In the case of time-space prediction, the
two-dimensional images of ${\cal {E}}$, i.e., ${\cal {E}}_w$, are
organized differently: the diagonal $n+\tau_w =1$ does not include
all trivial strategies, and the $(n,\tau_w)$ diagram  does not
include all optimal strategies (see Fig.~2).

Nevertheless, $n(\tau_w)$ is a convenient tool to visualize such
optimal strategies as are suitable for a trade-off between $n$ and
$\tau_w$. However, if $\{ w_i\} \ne \{ \lambda_i/\lambda \}$, then
the distance of $n(\tau_w)$ from the diagonal $n+\tau_w =1$ does not
tell us anything about the prediction potential of the relevant
strategies. To learn something about this potential, we need the
image of trivial strategies $D$ on the $(n,\tau_w)$ plane. The lower
boundary of $D_w$ may be very close to the ideal strategy with the
errors (0, 0).

Let us consider an example. The relative intensity (RI) method
(Tiampo et al., 2002) predicts the target event in that location
where the historical seismicity rate, $f(g)$, is the highest, $f>c$.
The RI is a typical example of a trivial strategy occasionally
employed as an alternative to meaningful prediction techniques (see,
e.g., Marzocchi et al., 2003). By the RI method, $\tau_i=1$, if
$f>c$ in the $i${\it -th} bin and $\tau_i=0$ otherwise. If $\{
w_i=\lambda_i/\lambda \}$, then

\begin{eqnarray*}
1 - n = \int_{f>c} f(g)\,dg = \tau_w  ,
\end{eqnarray*}
i.e., $n+\tau_w =1$ for any level $c$. If $w_i =\vert G_i \vert
/\vert G \vert $, where $\vert G \vert $ is the area of $G$, then
the curve $n(\tau_w)$ can be obtained by using (\ref{qq12}) (see
also Zechar and Jordan,2008). The curve passes close to (0,0), if
most of the target events occur in a relatively small area, say,
$\lambda_1/\lambda$ is close to 1 and $w_1$ is close to 0.

One gets a unique set of weights by choosing $w_i=\lambda_i/\lambda$
(see {\bf 3.2}, {\bf 3.8}). It is only in this particular case that
all trivial strategies are projected onto the diagonal
$\tilde{D}\,:\,n+\tau_w =1$, and $1-n(\tau_w)$ is a ROC curve.
Besides, the projection on the $(n,\tau_w)$ plane preserves the
relative distance between any strategy and the set of trivial
strategies . To be more specific, the following relations are true:

\begin{eqnarray}\label{qq25}
1 - n - \sum^k_{i=1} \tau_i \lambda_i/\lambda = \frac{\rho
(Q,D)}{\rho (O,D)} = \frac{\rho (Q_w,\tilde{D})}{\rho
(O_w,\tilde{D})} = 1 - n - \tau_w \label{qq25}
\end{eqnarray}
(Molchan and Keilis-Borok, 2008). Here, $\rho $ is the Euclidean
distance, e.g., $\rho (O,D)$ is the distance from $Q=(n,\btau)$ to
the hyperplane $D=\{ n + \sum \tau_i \lambda_i/\lambda =1\}$, and
$O=(0,0\ldots 0)$ corresponds to the ideal strategy. The right-hand
side of (\ref{qq25}) is known in the contingency table analysis as
the HK skill score (Hanssen-Kuiper, 1965). Consequently, when $\{
w_i=\lambda_i/\lambda \}$, the quantity
$H=\max\limits_{\tau_w}(1-n(\tau_w)-\tau_w)$ gives the greatest
relative distance between the optimal and the trivial strategies.

The choice of $\{ w_i\}$ at the research stage instead of $\{
\lambda_i/\lambda \}$ is justified by difficulties in the way of
estimating the $\{ \lambda_i \}$. This justification is illusory,
however. One must know the lower boundary of $D_w$ in order to
answer the question of how nontrivial the $n(\tau_w)$ diagram is.
But this again requires knowledge of the $\{ \lambda_i \}$ (see
(\ref{qq12}) and Fig.~2).

2. {\it The relation to the SDT}. In recent years the studies in
earthquake prediction are actively using the Signal Detection Theory
(SDT) developed in the late 1980s in the atmospheric sciences (see,
e.g., Jalliffe and Stephenson, 2003 and the references therein).The
main object of this theory is a warning system, which characterizes
the state of hazard by a scalar quantity $\xi $. The system is
tested by making $K\gg 1$ trials in which the $i${\it -th} event $\{
\xi >u\}$ is interpreted as an alarm, $\widehat{x}_i=1$, otherwise
$\widehat{x}_i=0$. The results are compared with observations $x =
\mbox {\rm Yes}$  or No with respect to a target event. Any
dependence between the members of the sequence $\{ (\widehat{x}_i,
x_i)\}$ is ignored a priori. It is required only that the rate of
target events ($x = \mbox {\rm Yes}$) should be $0<s<1$. This
condition is essential for getting an acceptable estimate for the
simultaneous distribution of $(\widehat{x}_i, x_i)$. Note that $s=0$
in our approach.

Two problems are formulated: assessing the prediction performance
and choosing the threshold u in a rational manner. The first problem
is attacked using the $2 \times 2$ contingency table of forecasts
and the second by using the ROC diagram related to the hypothesis
testing about the conditional distribution of $\xi$ given $x = \mbox
{\rm Yes}$ and given $x = \mbox {\rm No}$.

In our terminology this situation is one with discrete "time" where
the data $I$ in a trial are given by $\xi $. Therefore, the SDT is
equivalent to the analysis of the time prediction of earthquakes
using a specified  precursor/algorithm, even though the prediction
of large earthquakes  involves $s\ll 1$. The $\mbox {\rm ROC}/n(\tau
)$ diagram then quantifies the predictive potential of a precursor,
$\xi $ in this case. All meaningful strategies are functions of $\xi
$, hence reduce to choosing the level $u$.

In the  case of any data, $I(t),\, n(\tau)$ characterizes the
prediction performance of $\{ I(t)\}$ and gives the lower bound to
ROC curves for any algorithm based on $\{ I(t)\}$. The studies of
Molchan (1990, 1997) answer the question of how the quantity $\xi $
should be constructed for the original prediction data and why the
relation to hypothesis testing arises at all.

The gist of the matter lies in the fact that the $2\times 2$
contingency table is defined by three parameters $(n,\tau ,s)$, and
the program of prediction optimization is formulated, explicitly or
implicitly, in terms of that table. As a result, we have to deal
with local optimal strategies only. When real time is incorporated
in the SDT framework, there arise additional parameters that are
important for seismological practice, e.g., the rate of connected
alarms (alarm clusters) $\nu $. The optimization of the loss
function $\varphi =an+b\tau +c\nu $ at once gets us beyond the SDT
framework and its tools. The strategies that optimize $\varphi $ are
not locally optimal, and can be found from Bellman-type equations
(Molchan and Kagan, 1992; Molchan, 1997).

The use of the SDT approach in space-time prediction imposes a
rather unrealistic limitation: the spatial rate of target events
must be homogeneous. Otherwise,  the ROC diagram looses its meaning
and becomes a $(n,\tau_w)$ diagram (see Fig.~2).

\bigskip
\centerline {\it Acknowledgements}
\bigskip

This work was supported by the Russian Foundation for Basic Research
through grant 08-05-00215. I thank D.L. Turcotte for useful
discussions, which have stimulated the writing of the present paper.

\newpage

\centerline {R~{\footnotesize E~F~E~R~E~N~C~E~S}}

\bigskip

\begin{description}

\item
Bolshev,~L.N. and Smirnov,~V.N., Tables of mathematical statistics,
(Nauka, Moscow 1983).

\item
Harte,~D. and Vere-Jones,~D. (2005), The Entropy Score and Its Uses
in Earthquake Forecasting, Pure Appl. Geophys. 162, 1229-1253.

\item
Hanssen,~A.W. and Kuipers,~W.J.A. (1965),  On the relationship
between the frequency of rain and various meteorological parameters.
Modedeelingen en Verhandelingen, Royal Notherlands Meteorological
Institute, 81

\item
Jolliffe,~I.T. and Stephenson,~D.B. (eds.), Forecast Verification: a
Practitioner's Guide in Atmospheric Science (John Wiley \& Sons,
Hoboken 2003).

\item
Kagan,~Y.Y. (2007), On Earthquake Predictability Measurement:
Information Score and Error Diagram, Pure Appl. Geophys. 164, 1947-
1962.

\item
Keilis-Borok,~V.I. and Soloviev,~A.A. (eds.),  Nonlinear Dynamics of
the Lithosphere and Earthquake Prediction (Springer-Verlag,
Berlin-Heidelberg 2003).

\item
Keilis-Borok,~V.I., Shebalin,~P., Gabrielov,~A., Turcotte,~D.
(2004). Reverse Tracing of Short-term Earthquake Precursors, Phys.
Earth.  Planet.  Inter. 145, 75-85.

\item
Kossobokov,~V.G. (2005), Earthquake Prediction: Principles,
Implementation, Perspectives, Computational Seismology, Iss.  36-1,
3-175, (GEOS, Moscow).

\item
Lehmann,~E.L., Testing Statistical Hypotheses (J.~Wiley\& Sons. New
York 1959).

\item
Marzocchi,~W., Sandri,~L., and Boschi,~E.(2003), On the Validation
of   Earthquake-forecasting Models: the Case of Pattern Recognition
Algorithms, Bull. Seism. Soc. Am. 93, 5, 1994-2004.

\item
Molchan,~G.M. (1990), Strategies in strong earthquake prediction,
Phys.  Earth  Planet. Inter. 61(1-2), 84-98

\item
Molchan,~G.M. (1991), Structure of Optimal Strategies of Earthquake
Prediction Tectonophysics 193, 267-276.

\item
Molchan,~G.M. (1997), Earthquake Prediction as a Decision Making
Problem, Pure Appl. Geophys. 149, 233-247.

\item
Molchan,~G.M., Earthquake Prediction Strategies: a Theoretical
Analysis. In Nonlinear dynamics of the Lithosphere and Earthquake
Prediction (eds. Keilis-Borok,~V.I. and Soloviev,~A.A.)
(Springer-Verlag, Berlin-Heidelberg 2003), pp.209-237.

\item
Molchan,~G.M. and Kagan,~Y.Y. (1992), Earthquake Prediction and its
Optimization, J. Geophys. Res. 97, 4823-4838.

\item
Molchan,~G.M. and Keilis-Borok,~V.I., (2008), Earthquake Prediction:
Probabilistic Aspect, Geophys. J. Int. 173, 1012-1017.

\item
Shcherbakov,~R., Turcotte,~D.L., Holliday,~J.R., Tiampo,~K.F., and
Rundle,~J.B. (2008), A Method for Forecasting the   Locations of
Future Large Earthquakes: An Analysis and Verification, Geophys.
Res. Lett., DOI: 1010291 (in press).

\item
Shen,~Z.-K., Jackson,~D.D., and Kagan,~Y.Y. (2007),  Implications of
Geodetic Strain Rate for Future Earthquakes, with a Five-Year
Forecast of M5 Earthquakes in Southern California, Seismol. Res.
Lett. 78(1), 116-120.

\item
Swets,~J.A. (1973), The Relative Operating Characteristic in
Psychology, Science 182, 4116, 990-1000.

\item
Tiampo,~K.F., Rundle,~J.B., McGinnis,~S., Gross,~S., and Klein,~W.
(2002), Mean Field Threshold Systems and Phase Dynamics:  An
Application to Earthquake Fault Systems, Europhys. Lett. 60(3),
481-487.

\item
Zechar,~J.D. and Jordan,~Th.,H. (2008), {\it Testing alarm-based
earthquake predictions}, Geophys. J. Int. 172, 715-724

\end{description}

\newpage

\centerline{{\bf Figure captions}}

\bigskip

\noindent Fig.~1. Space-time prediction characteristics: $n$ vs.
$\btau =(\tau_1, \ldots ,\tau_k)$ (the horizontal axis is
multidimensional)

\noindent {\it Notation}: ${\cal {E}}(I)$ represents all strategies
based on the data $I$, the hyperplane $D$ represents the trivial
strategies (random guesses), and the surface $n(\btau )$ the optimal
strategies (the error diagram). The level sets of the loss function
$\varphi (n,\btau )$ are shown by dashed lines, the characteristic
of the optimal prediction is a tangent point $Q$ between $n(\btau )$
and the suitable level set of $\varphi $.

Fig.~2. The reduced error diagram: $n$ vs. $\tau_w = \sum^k_{i=1}
\tau_i w_i$

\noindent {\it Notation}: ${\cal {E}}_w$ contoured by bold lines
represents all strategies ${\cal {E}}$ in the $(n,\tau_w)$
coordinates; the stippled zone $D_w$ represents the trivial
strategies; the broken line within $D_w$ illustrates the method used
to construct $D_w$, see {\bf 3.3}; the filled zone is the image of
the $n(\btau )$ diagram; isolines of the loss function $\varphi
=\psi (n,\tau_w)$ are shown by dashed lines; $\varphi $ yields the
optimal characteristics $Q_w=(n,\tau_w)$.

\newpage

\begin{figure} [p]
\centering{\includegraphics{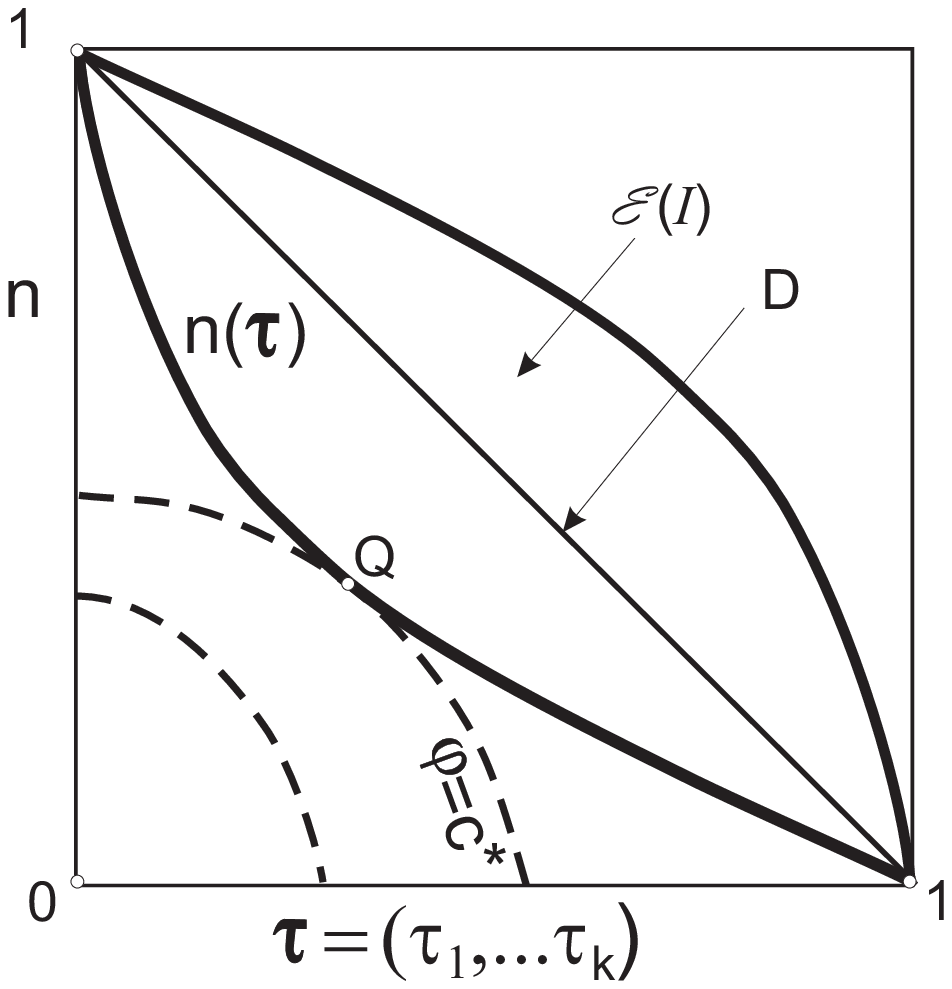}}

\end{figure}

\newpage

\begin{figure} [p]
\centering{\includegraphics{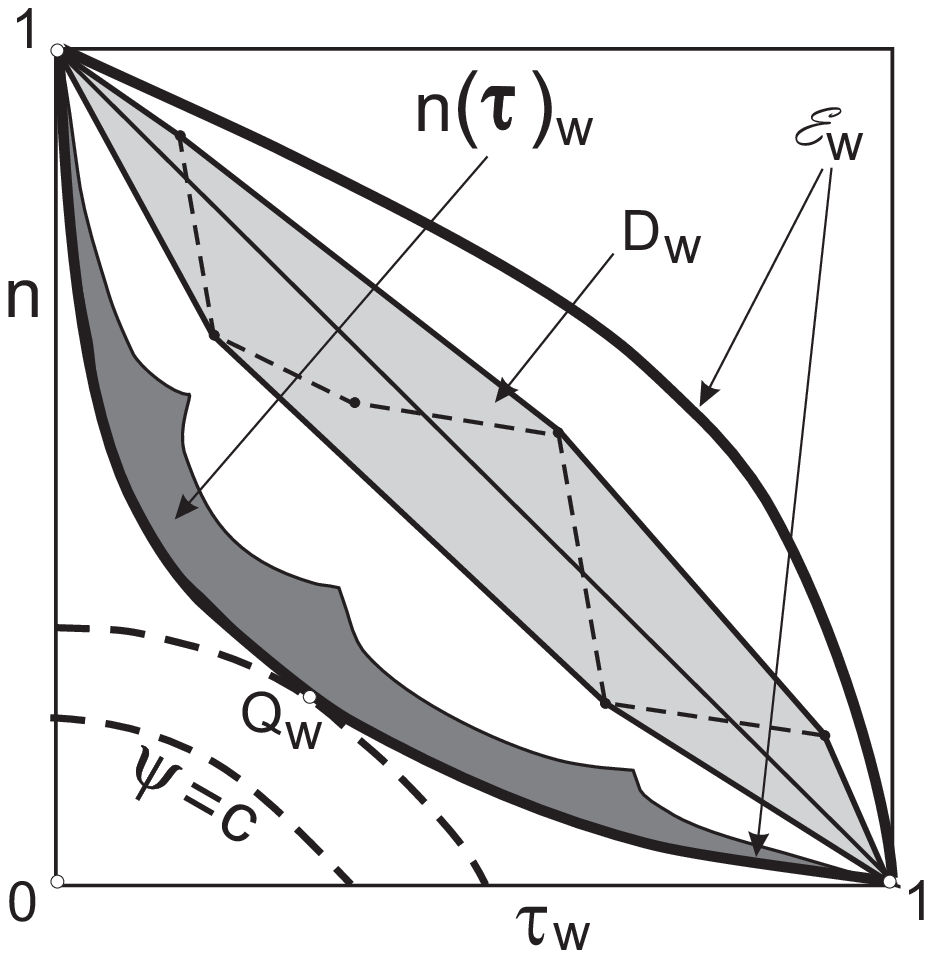}}

\end{figure}

\end{document}